\documentclass{article}

\usepackage[preprint]{neurips_2026}

\usepackage[utf8]{inputenc} % allow utf-8 input
\usepackage[T1]{fontenc}    % use 8-bit T1 fonts
\usepackage{hyperref}       % hyperlinks
\usepackage{url}            % simple URL typesetting
\usepackage{booktabs}       % professional-quality tables
\usepackage{amsfonts}       % blackboard math symbols
\usepackage{nicefrac}       % compact symbols for 1/2, etc.
\usepackage{microtype}      % microtypography
\usepackage{xcolor}         % colors
\usepackage{tabularx, tabulary}
\usepackage{multirow}
\usepackage[table]{xcolor}
\usepackage{tikz}
\usepackage{pgfplots}
\pgfplotsset{compat=1.18}
\usepackage{pgfplotstable}
\usepackage{subcaption}
\usepackage{wrapfig}
\usepackage{graphicx}
\usepackage{enumitem}
\usepackage{amsmath}
\usepackage{cleveref}

\newcommand{\phm}[1]{\vspace{.4em} \noindent\textbf{#1}\hspace{.5em}}
\newcommand{\sysname}{\emph{AB-Sparse}}
\renewcommand{\thefootnote}{\fnsymbol{footnote}}

\newcommand*\circleb[1]{\tikz[baseline=(char.base)]{
    \node[shape=circle,draw,inner sep=.5pt](char){\small #1};}}
\title{\sysname{}: Sparse Attention with Adaptive Block Size for Accurate and Efficient Long-Context Inference}

\author{%
    Di Liu$^{1}$\footnotemark[1], 
    Ruitian Wang$^{1}$\footnotemark[1], 
    Chen Chen$^{1}$\footnotemark[3], 
    Mingliang Gong$^{2}$\footnotemark[3]\\
    \textbf{
        Yongjie Yuan$^{2}$,
        Han Zhao$^{1}$ ,
        Yu Feng$^{1}$, 
        Quan Chen$^{1}$, 
        Minyi Guo$^{1}$
    } \\
    $^{1}$Shanghai Jiao Tong University \quad
    $^{2}$Ant Group \\
}

\begin{document}

\footnotetext[1]{Equal contribution.}
\footnotetext[3]{Corresponding authors: Chen Chen and Mingliang Gong.}
\renewcommand{\thefootnote}{\arabic{footnote}} 

\maketitle

\begin{abstract} 
As large language models scale to longer contexts, loading the growing KV cache during attention computation becomes a critical bottleneck.
Previous work has shown that attention computation is dominated by a small subset of tokens. 
This motivates block sparse attention methods that partition the KV cache into fixed-size blocks and selectively compute attention over those blocks exhibiting high importance.
However, these methods assign a uniform block size across all attention heads, implicitly assuming homogeneous behavior throughout the model. 
Our analysis reveals that this assumption is flawed: attention heads exhibit widely varying sensitivity to block granularity, and uniformity leads to suboptimal accuracy.
We present \sysname{}, a training-free algorithm-system co-designed framework that improves accuracy while preserving throughput. 
\sysname{} introduces lightweight adaptive block size allocation across attention heads to improve accuracy. 
To compensate for the additional memory overhead, it further employs lossless block centroid quantization. 
In addition, custom GPU kernels are developed to support efficient execution with variable block sizes.
Evaluation results demonstrate that \sysname{} achieves an accuracy improvement of up to 5.43 \% over existing block sparse attention baselines without throughput overhead.
\end{abstract}

\section{Introduction}

Large language models (LLMs)~\cite{brown2020language,liu2024deepseek,yang2025qwen3} are increasingly deployed in applications that demand long-context understanding, ranging from multi-document summarization~\cite{yang2025curiousllm} to repository-level code analysis~\cite{luo2024repoagent} and long-form reasoning~\cite{wei2022chain}.
While larger context windows enable these capabilities, they introduce significant challenges for efficient model serving.
At every decoding step, loading the entire KV cache from memory becomes a bottleneck that scales linearly with context length. 
For instance, the KV cache for a single request of 128K context length on Llama-3.1-8B~\cite{llama-3.1-8B} reaches 16GB, comparable to the model weights in size.

\begin{figure}[t]
    \centering
    \begin{minipage}[t]{0.48\textwidth}
        \centering
        \includegraphics[width=0.6\linewidth]{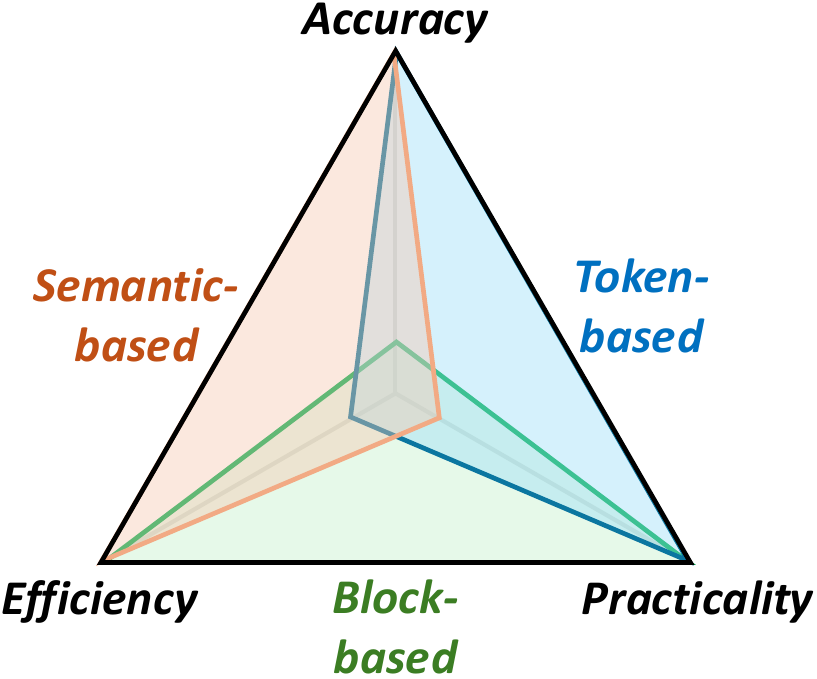}
        \caption{Qualitative comparison of various sparse attention paradigms.}
        \label{fig:bg_bottleneck}
    \end{minipage}
    \hfill
    \begin{minipage}[t]{0.48\textwidth}
        \centering
        \includegraphics[width=1\linewidth]{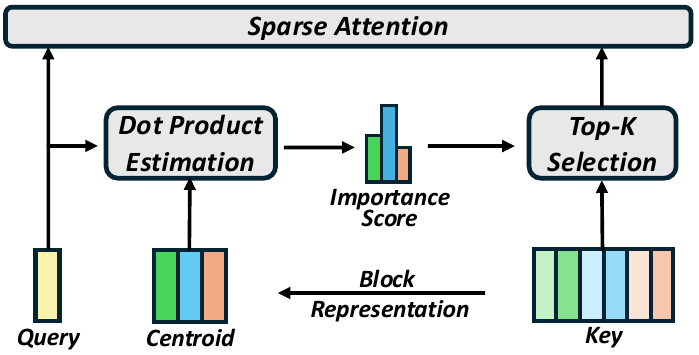}
        \caption{Illustration of block sparse attention workflow.}
        \label{fig:bg_block_sparse}
    \end{minipage}
\end{figure}

The key to addressing this bottleneck lies in the inherent sparsity of attention: only a small subset of tokens dominates the attention output~\cite{xiao2023efficient}. 
This property has inspired a growing body of sparse attention methods that can be categorized into three paradigms, as illustrated in~\Cref{fig:bg_bottleneck}. 
While all three paradigms reduce KV cache loading to lower inference latency, they each face different trade-offs among efficiency, practicality, and accuracy.
\textbf{Token-based} methods such as H\textsubscript{2}O~\cite{zhang2023h2o} and InfiniGen~\cite{lee2024infinigen} estimate per-token importance at every decoding step and select the most relevant tokens, but incur high per-step selection overhead, compromising efficiency.
\textbf{Semantic-based} methods such as Clusterkv~\cite{liu2024clusterkv} and RetroInfer~\cite{chen2025retroinfer} cluster tokens by key similarity and retrieve only the relevant clusters, but restructure the KV cache layout, compromising practicality with standard paged KV cache management~\cite{kwon2023efficient,zheng2024sglang,ye2025flashinfer}.
\textbf{Block-based} methods such as Quest~\cite{tang2024quest} and ArkVale~\cite{chen2024arkvale} partition the KV cache into fixed-size blocks and load only the Top-$K$ for attention computation, preserving both efficiency and practicality.
This makes block-based methods a promising foundation, and the key to fully unlocking their potential lies in improving accuracy without sacrificing throughput.

Our analysis reveals that the accuracy limitation of block-based methods stems from a fundamental yet overlooked assumption: a uniform block size is applied across all attention heads~\cite{tang2024quest,chen2024arkvale}.
However, attention heads are known to exhibit highly heterogeneous behaviors~\cite{xiao2024duoattention,wu2024retrieval}.
As shown in~\Cref{fig:mot_recall_curves}, our measurement study further reveals that this heterogeneity extends to their sensitivity to block granularity, with heads varying significantly in their block size preference.
Forcing a uniform block size across all heads thus creates an inherent tension: for heads that require fine-grained resolution, an overly large block size coarsens selection granularity, causing critical tokens to be missed; for heads that are insensitive to granularity, an overly small block size unnecessarily increases the number of blocks, amplifying computation and memory cost.

The attention head heterogeneity necessitates adaptive block size allocation: assigning finer granularity to sensitive heads to preserve accuracy, while allowing coarser blocks for insensitive heads to reduce overhead. 
However, realizing this in practical inference systems is non-trivial, with challenges arising from three aspects.
First, adjusting block sizes at runtime requires recomputing centroids for all blocks, which is prohibitively expensive; a lightweight mechanism is needed to determine per-head block size assignments prior to deployment.
Second, assigning smaller blocks to sensitive heads multiplies their centroid count, introducing significant memory overhead that calls for a lossless compression scheme to reduce centroid footprint.
Third, heterogeneous block sizes lead to non-uniform centroid counts across heads and conflict with the uniform page size assumption in existing paged KV cache systems, necessitating custom GPU kernels for efficient execution.

We present \sysname{}, a training-free algorithm-system co-designed framework that addresses these challenges through three tightly integrated components.
First, \sysname{} introduces a lightweight calibration-driven profiling strategy; it exploits the stability of per-head block size sensitivity across diverse inputs to derive reliable assignments prior to deployment.
Second, observing that block centroids are precision-insensitive as they are used only for ranking rather than for attention computation, \sysname{} proposes lossless centroid quantization to reduce memory footprint.
Third, \sysname{} implements custom GPU kernels to support efficient execution with adaptive block sizes. An indexing mechanism enables variable-length batched execution across heads, while a page mapping mechanism maintains compatibility with standard paged KV cache management.

We evaluate \sysname{} on three widely used open source models across two long-context benchmarks. \sysname{} consistently outperforms existing block sparse attention baselines, achieving up to 5.43\% accuracy improvement without sacrificing throughput. Our contributions are summarized as follows:
\begin{itemize}[leftmargin=1.5em]
    \item We conduct a systematic measurement study on block size sensitivity, revealing that attention heads exhibit substantial heterogeneity in block granularity preference (\S\ref{sec:moti_empirical}).
    \item We propose an adaptive block size allocation strategy based on lightweight calibration-driven profiling. We introduce lossless centroid quantization to reduce memory footprint. We design custom GPU kernels with two key mechanisms: indexing for variable-length batched execution, and page mapping for compatibility with standard paged KV cache management (\S\ref{sec:design}).
    \item We evaluate \sysname{} on three LLMs across two long-context benchmarks, demonstrating consistent accuracy improvements of up to 5.43\% over uniform-block-size baselines without throughput overhead (\S\ref{sec:eval}).
\end{itemize}

\section{Background and Motivation}

\subsection{LLMs and Attention Operation} 
The core component of LLMs is the attention operation~\cite{vaswani2017attention}.
At each decoding step $t$, the attention operation computes the dot product between the query vector $\mathbf{q}_t \in \mathbb{R}^{1\times d}$ (where $d$ is the hidden dimension) and the key vectors of all preceding tokens $\mathbf{k}_i \in \mathbb{R}^{1\times d}$ (for $i\leq t$). 
This product is scaled by $d^{-\frac{1}{2}}$ and normalized through Softmax function to yield the attention score $a_{t,i}$. 
These scores then weight the value vectors $\mathbf{v}_i$, resulting in the final attention output $\mathbf{o}_t$.
\begin{equation}
    z_{t,i} = \frac{\mathbf{q}_{t} \cdot \mathbf{k}^T_i}{\sqrt{d}}, \quad
    a_{t,i} = \dfrac{
        e^{z_{t,i}}
    }{
        \sum_{j=1..t}{e^{z_{t,j}}}
    }, \quad
    \mathbf{o}_t = \sum_{i=1..t} a_{t,i}\cdot \mathbf{v}_i
    % \label{eq:full-attention}
    \label{eq:attn}
\end{equation}

The attention module is typically composed of multiple components, each referred to as an attention head~\cite{vaswani2017attention,Ainslie2023gqa}. 
Each head independently performs computation as in~\Cref{eq:attn} and captures diverse features from different subspaces. 
The results from all heads are then aggregated to yield the output.

LLM inference consists of two stages: the prefill phase and the decoding phase.
The prefill phase processes all prompt tokens simultaneously with $O(n^2)$ complexity.
In the decoding phase, each newly generated token attends to all preceding tokens. 
A standard optimization is to cache these KV states (KV cache), reducing the complexity to $O(n)$.
However, loading the full KV cache becomes a bottleneck as context length grows, making the decoding phase memory-bound.

\subsection{Block Sparse Attention}
A promising approach to reducing KV cache loading is to exploit the inherent sparsity of attention, where only a small subset of tokens dominates the output~\cite{jiang2024minference,liu2024retrievalattention,chen2024magicpig}.
Among various sparse attention methods, block-based approaches such as Quest~\cite{tang2024quest} and ArkVale~\cite{chen2024arkvale} have gained widespread adoption due to their efficiency and practicality with standard paged KV cache layouts~\cite{kwon2023efficient,zheng2024sglang,ye2025flashinfer}.

\Cref{fig:bg_block_sparse} illustrates the common workflow of block sparse attention.
The KV cache is partitioned into blocks of equal size $B$, where each block is represented by a centroid $c_i$\footnote{Various methods have been proposed to compute block centroids, such as mean pooling~\cite{lu2025moba} and max-min pooling~\cite{tang2024quest}. The block representation strategy is orthogonal to this work.}. 
During the estimation stage, the query vector $q_t$ computes the dot product with all centroids, yielding importance score $r_{t,i} = q_t \cdot c_i^\top$.
The Top-$K$ blocks with the highest scores are then selected for approximate attention.

The choice of block size $B$ governs a fundamental trade-off between accuracy and efficiency.
A larger $B$ coarsens the centroid representation, degrading Top-$K$ selection accuracy. Conversely, a smaller $B$ increases the total centroid count, amplifying both memory and computation cost.

\subsection{Adaptive Block Size Allocation}
\label{sec:moti_empirical}

Existing block sparse attention methods fix $B_h = B$ for all heads.
However, attention heads exhibit substantial heterogeneity in how critical tokens are distributed across the KV cache:
for some heads critical tokens are densely clustered, while for others they are sparsely scattered.
This heterogeneity leads to varying sensitivity across heads, rendering a uniform block size inherently suboptimal.

We conduct a systematic analysis of per-head block size sensitivity on Llama-3.1-8B~\cite{llama-3.1-8B} and Qwen3-8B~\cite{qwen3} using the Wikipedia dataset~\cite{wikipedia} with context length of 32K tokens.
For each attention head, we vary the block size over $\{16, 32, 64\}$ while maintaining a fixed token budget of 4096 \footnote{Top-$K$ blocks are chosen such that the total tokens across selected blocks equals 4096. This trend is consistent across different token budgets.}.
We measure attention recall—the fraction of total attention score attributed to the tokens in the selected blocks—as a direct indicator of block selection quality.

\begin{figure}[t]
    \centering
    \begin{minipage}[t]{0.48\linewidth}
        \centering
        \includegraphics[width=0.95\linewidth]{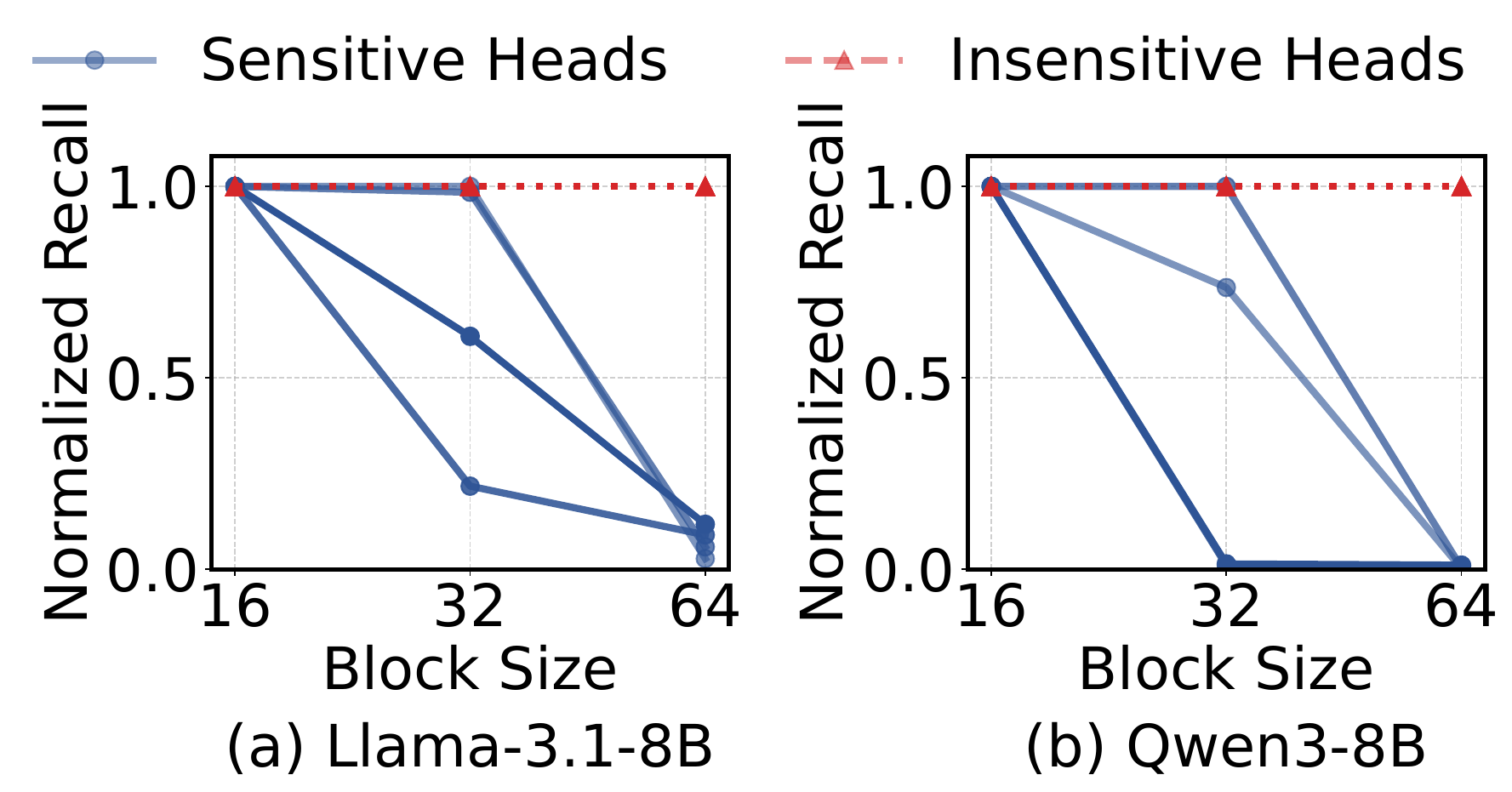}
        \caption{
            Normalized recall curves across block sizes, where normalization is performed with respect to the recall at block size 16. Insensitive heads maintain near-perfect normalized recall across all block sizes, while sensitive heads degrade sharply as block size increases.
        }
        \label{fig:mot_recall_curves}
    \end{minipage}
    \hfill
    \begin{minipage}[t]{0.48\linewidth}
        \centering
        \includegraphics[width=\linewidth]{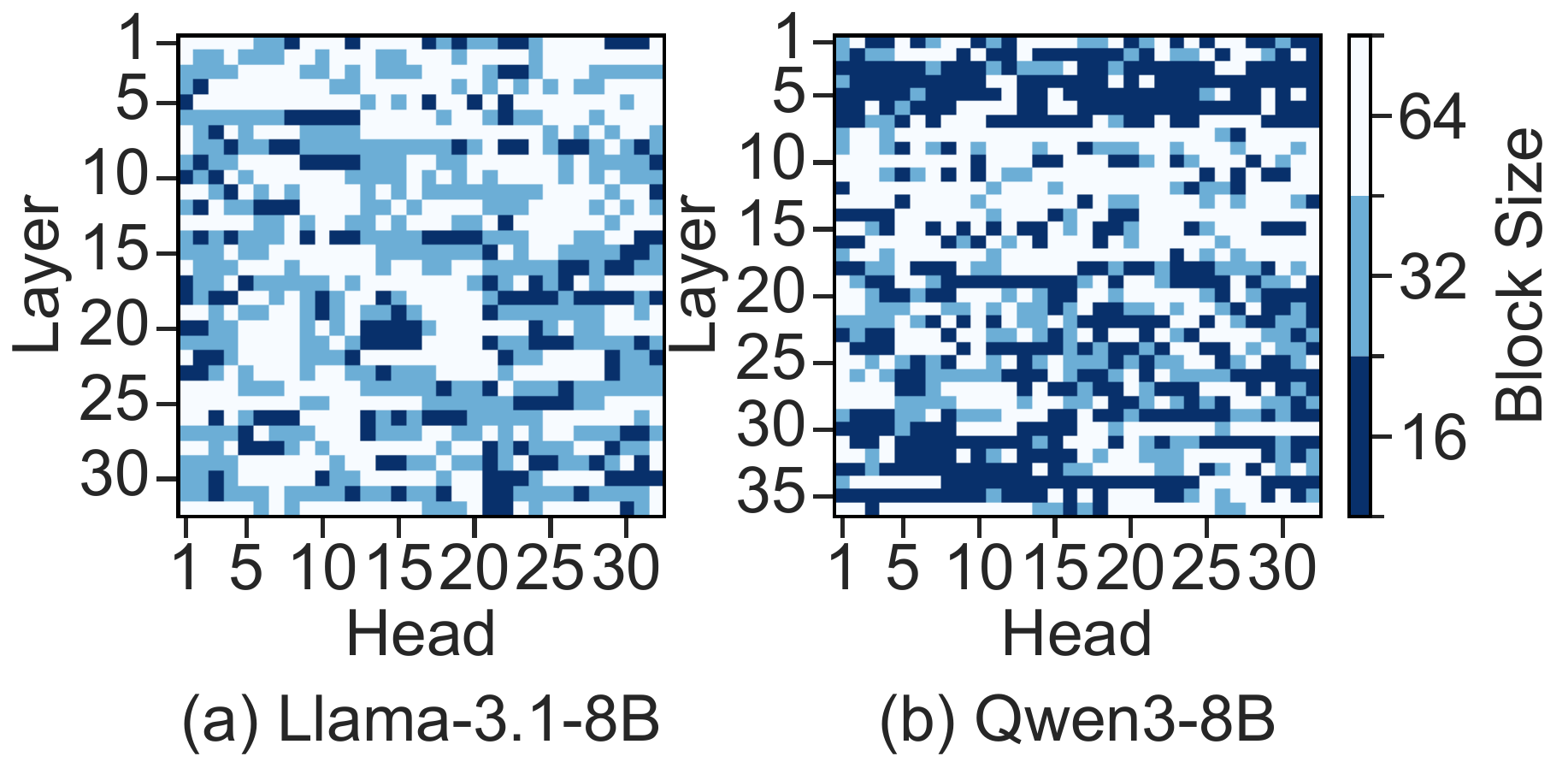}
        \caption{
            Heatmap of the minimum block size required to retain 98\% of peak recall for each attention head across layers. The wide variation across heads and layers indicates that no single uniform block size is simultaneously efficient and accurate.
        }
        \label{fig:mot_block_size_dist}
    \end{minipage}
\end{figure}

\phm{Attention heads exhibit heterogeneous block size sensitivity.}\Cref{fig:mot_recall_curves} shows the normalized recall curves of representative attention heads for both models.
Insensitive heads maintain near-perfect recall across all block sizes, while sensitive heads degrade sharply, dropping below 0.1 at block sizes as small as 32.
This reveals that attention heads vary substantially in their sensitivity to block granularity, exhibiting distinct block-size preference.

\phm{Adaptive allocation outperforms uniform block sizes.}\Cref{fig:mot_block_size_dist} shows that the minimum block size to retain 98\% of peak recall (i.e., recall at the smallest block size) varies widely across heads and layers, implying that no single uniform block size is simultaneously efficient and accurate.
For instance, under a uniform block size of 32, the average recall is only 89.7\% and 77.8\% on Llama-3.1-8B and Qwen3-8B, whereas adaptive allocation achieves 98\% with a larger average block size of 44.2 and 39.5.
This demonstrates that adaptive per-head block size allocation has the potential to improve recall without reducing the average block size.

These findings motivate the design of \sysname{}, which adaptively assigns per-head block sizes to improve accuracy while maintaining system efficiency.

\section{\sysname{} Design}
\label{sec:design}

Our empirical findings in~\Cref{sec:moti_empirical} uncover substantial heterogeneity in block size sensitivity across attention heads, which has been overlooked by existing block sparse attention methods. 
This highlights the potential for adaptive per-head block size allocation to improve accuracy without sacrificing system efficiency. 
Building on this insight, we first outline the key challenges and our system architecture in \S\ref{sec:overview}, then detail each component in the following subsections.

\subsection{Overview}
\label{sec:overview}

\begin{wrapfigure}{r}{0.5\linewidth}
    \centering
    \includegraphics[width=\linewidth]{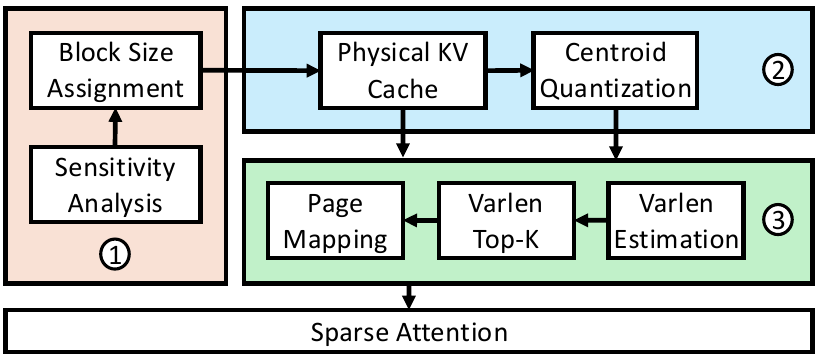}
    \caption{Architecture of \sysname{}.}
    \label{fig:arch}
\end{wrapfigure}

Adaptive block size allocation entails design challenges in three aspects of the practical inference system. 
First, adaptivity requires a block size assignment for each attention head; dynamically adjusting assignments at runtime is prohibitively expensive, as it requires recomputing centroids over all key vectors. 
Second, assigning smaller blocks to sensitive heads significantly increases the number of centroids that must be stored; as context length grows, this overhead scales linearly with sequence length, threatening to bottleneck decoding throughput. 
Third, heterogeneous block sizes across heads break the execution uniformity assumed by standard batched kernels and are incompatible with existing inference systems that universally adopt fixed-size paged KV cache management.

\sysname{} addresses these challenges with three tightly integrated designs, as summarized in~\Cref{fig:arch}. 
\circleb{1}: Observing that per-head block size sensitivity remains stable across diverse inputs, \sysname{} profiles recall sensitivity on a small calibration set to derive reliable per-head block size assignments (\S\ref{sec:design_alloc}). 
\circleb{2}: Recognizing that block centroids are precision-insensitive as they serve solely for ranking rather than attention computation, \sysname{} applies lossless centroid quantization to reduce memory footprint without degrading block selection accuracy (\S\ref{sec:design_quant}).
\circleb{3}: \sysname{} implements dedicated GPU kernels with an indexing mechanism for variable-length batched execution and a page mapping mechanism for compatibility with standard paged KV cache management (\S\ref{sec:design_kernel}).

\subsection{Lightweight calibration-driven profiling}
\label{sec:design_alloc}

% \begin{wrapfigure}{t}{0.35\linewidth}
%     \centering
%     \includegraphics[width=1\linewidth]{figure/solu_genera.pdf}
%     \caption{Recall comparison between adaptive and uniform block size. The adaptive assignments are calibrated solely on wikipedia~\cite{wikipedia}. Despite this, they consistently outperform uniform block size across all four RULER~\cite{hsieh2024ruler} tasks.}
%     \label{fig:solu_genera}
% \end{wrapfigure}

\begin{figure}[t]
    \centering
    \begin{minipage}[t]{0.48\linewidth}
        \centering
        \includegraphics[width=0.95\linewidth]{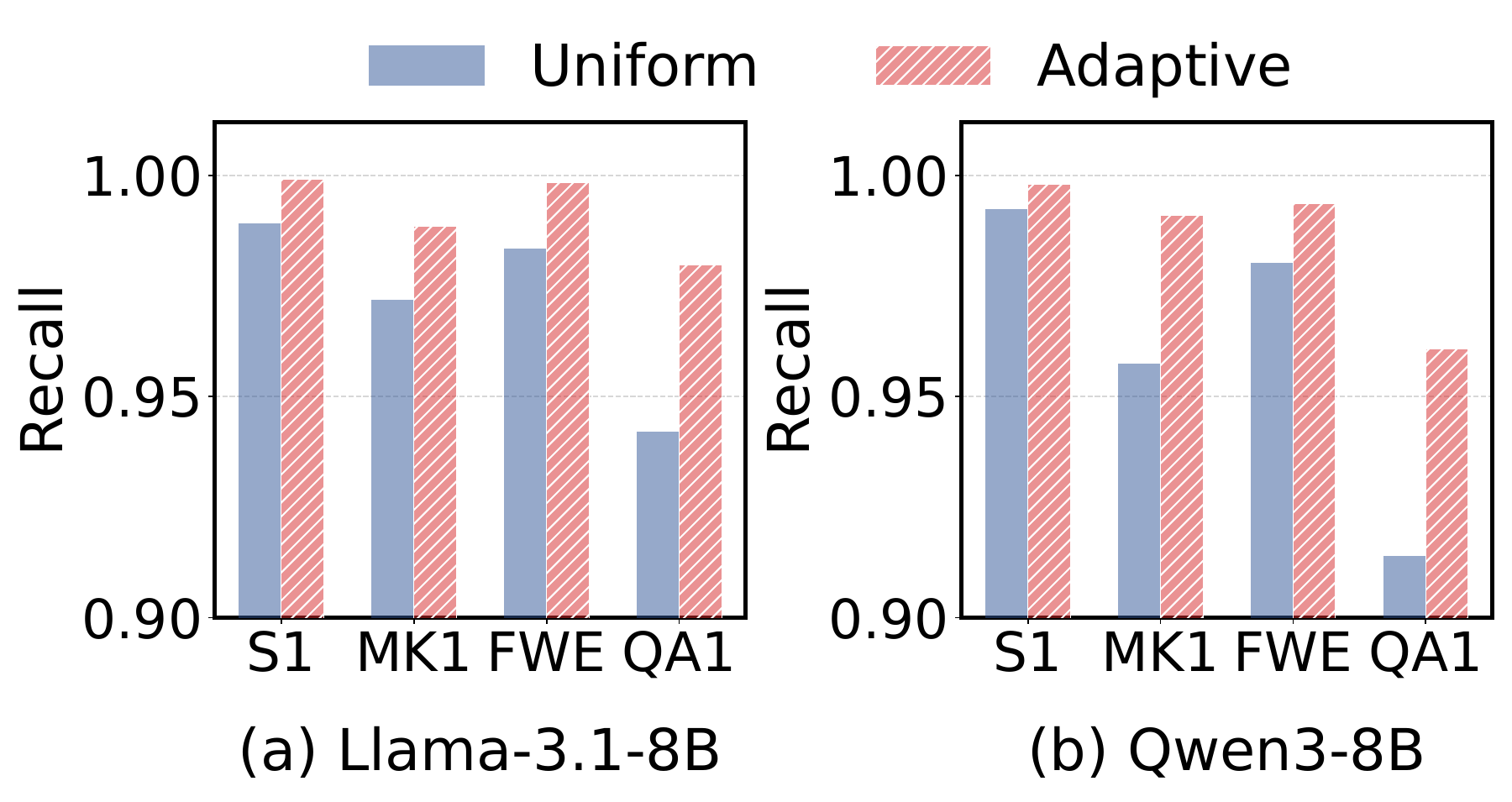}
        \caption{
            Recall comparison between adaptive and uniform block size. The adaptive assignments are calibrated solely on wikipedia~\cite{wikipedia}. Despite this, they consistently outperform uniform block size across all RULER~\cite{hsieh2024ruler} tasks.
        }
        \label{fig:solu_genera}
    \end{minipage}
    \hfill
    \begin{minipage}[t]{0.48\linewidth}
        \centering
        \includegraphics[width=0.9\linewidth]{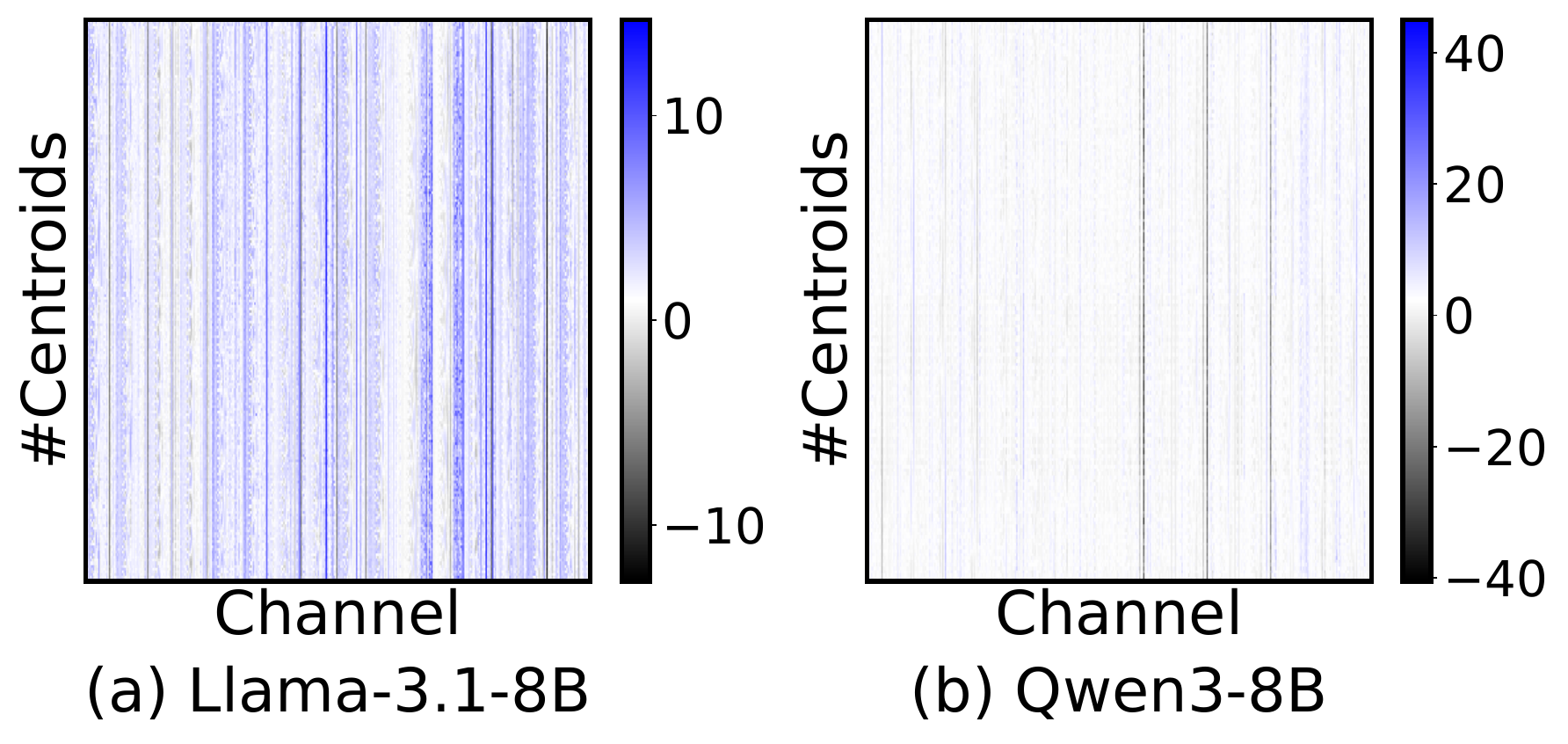}
        \caption{
            Centroid value distribution of Llama-3.1-8B and Qwen3-8B. The column-wise patterns indicate that centroid values are tightly clustered per channel, supporting the use of per-channel quantization.
        }
        \label{fig:solu_quant_channel}
    \end{minipage}
\end{figure}

Determining per-head block size assignments is non-trivial. Adjusting block sizes dynamically requires recomputing centroids over the entire KV cache under each candidate block size, whose cost scales linearly with context length and is prohibitively expensive at inference time.

The key insight is that per-head block size sensitivity is stable across diverse inputs.
Previous work has shown that individual attention heads learn specialized roles, such as local pattern matching and long-range retrieval~\cite{xiao2024duoattention,wu2024retrieval,jiang2024minference}.
These roles are determined by learned parameters and thus remain consistent across inputs.
Our finding that block size preference is similarly head-specific and input-invariant aligns with this understanding.
Heads that are sensitive to block size remain sensitive regardless of the input, and vice versa. 
This suggests that a one-time offline calibration is sufficient to derive reliable assignments that generalize across requests.

Concretely, \sysname{} evaluates attention recall on 50 calibration samples from wikipedia~\cite{wikipedia}. 
For each head, the largest block size that satisfies a recall retention threshold $\tau$ is selected:
\begin{equation}
    B_h^* = \max \{ B \mid \text{Recall}(h, B) \geq \tau \cdot \text{Recall}(h, B_{\min}) \}
    \label{eq:calib}
\end{equation}
where $B_{\min}$ is the smallest candidate block size, and $\tau$ serves as a knob to balance recall preservation and centroid overhead.

To validate generalization, we evaluate the derived assignments on four tasks from RULER~\cite{hsieh2024ruler}, covering diverse long-context scenarios. As shown in~\Cref{fig:solu_genera}, despite being calibrated solely on wikipedia, the assignments consistently outperform uniform block size across all tasks and models with a comparable average block size. This confirms that per-head block size sensitivity is stable across tasks, and that a one-time calibration is sufficient for reliable deployment.

\subsection{Lossless centroid quantization}
\label{sec:design_quant}

Adaptive block size allocation assigns smaller blocks to sensitive heads, which can significantly increase their centroid count and amplify memory overhead. 
To keep this overhead bounded, centroid compression is necessary.
We observe that centroid vectors are used solely for ranking and selecting the Top-$K$ blocks, rather than directly contributing to attention outputs. 
This precision-insensitive property makes quantization a natural fit for centroid compression.
However, naively reducing bit width to very low precision risks degrading block selection accuracy. This necessitates a quantization scheme that maximizes compression while preserving accuracy.

A closer examination reveals that for each position along the head dimension (i.e., each channel), centroid values across different blocks follow a concentrated distribution. 
As shown in~\Cref{fig:solu_quant_channel}, centroid values exhibit clear column-wise patterns across both models, confirming that values within each channel are tightly clustered. 
This intra-channel similarity makes a single scaling factor per position sufficient to capture the value range without introducing large quantization error, enabling more aggressive compression while preserving ranking fidelity.

\begin{figure}[t]
    \centering
    \begin{minipage}[t]{0.48\textwidth}
        \centering
        \includegraphics[width=\linewidth]{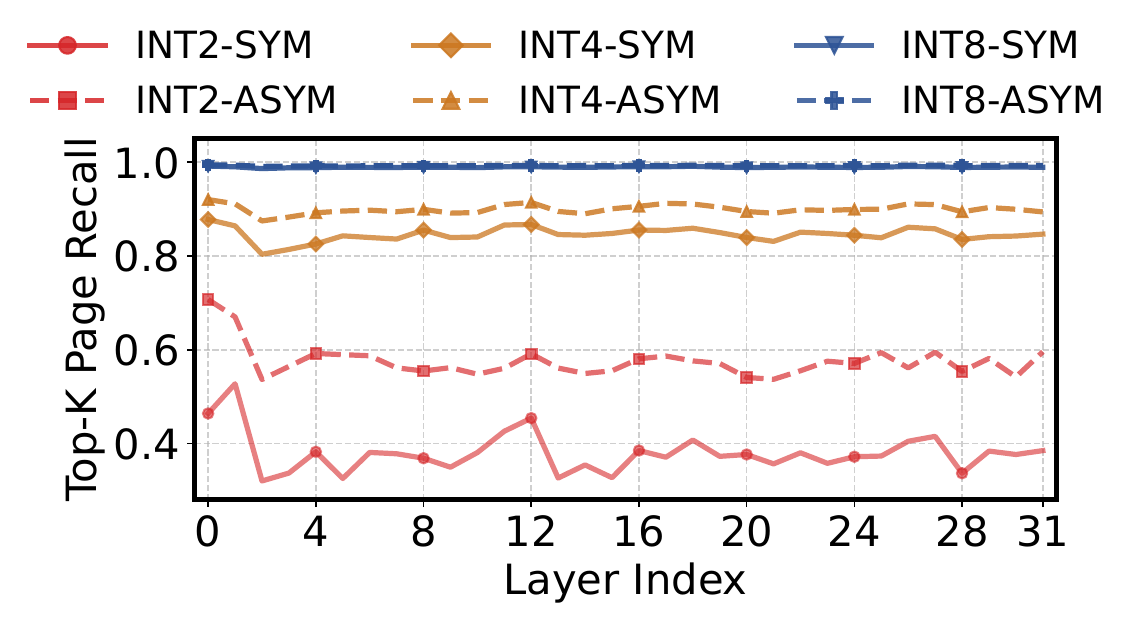}
        \caption{Top-$K$ page recall across layers on Llama-3.1-8B under different quantization bit widths and strategies. INT4 asymmetric per-channel quantization consistently maintains recall above 0.9 across all layers.}
        \label{fig:solu_quant}
    \end{minipage}
    \hfill
    \begin{minipage}[t]{0.48\textwidth}
        \centering
        \includegraphics[width=0.98\linewidth]{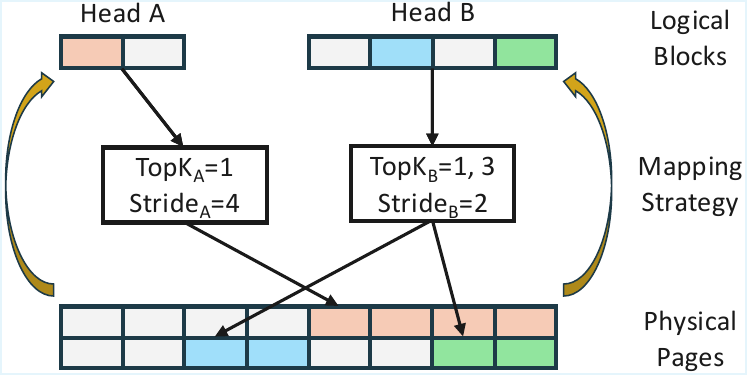}
        \caption{Illustration of the page mapping process. Logical blocks of varying sizes are mapped to contiguous physical pages via a block-to-page stride, enabling variable block size to interface with standard paged KV cache management.}
        \label{fig:solu_mapping}
    \end{minipage}
\end{figure}

To identify the optimal quantization scheme, we measure Top-$K$ page recall across layers on Llama-3.1-8B under different bit widths (INT2, INT4, INT8) and quantization strategies (symmetric and asymmetric\footnote{Symmetric quantization maps values to a zero-centered range, while asymmetric quantization additionally uses a zero-point offset to handle skewed distributions.}). 
As shown in \Cref{fig:solu_quant}, lower bit widths (INT2) suffer significant recall degradation across layers. 
While INT4 symmetric quantization improves over INT2, it still fails to consistently maintain high recall. 
INT4 asymmetric per-channel quantization, on the other hand, achieves recall above 0.9 across all layers and both models. 
Although INT8 quantization yields slightly higher recall, INT4 asymmetric strikes a better balance between accuracy and memory efficiency. 
\sysname{} therefore adopts INT4 asymmetric per-channel quantization.

\subsection{Efficient custom GPU kernels}
\label{sec:design_kernel}

Modern GPU kernels achieve high throughput by batching all attention heads into a single kernel launch, which requires each head to have the same number of centroids for aligned execution. 
In addition, existing inference systems manage the KV cache in fixed-size physical pages, assuming a uniform block-to-page mapping across all heads.
Heterogeneous block sizes break both assumptions.
Different heads have varying centroid counts, forcing standard batched execution to resort to either wasteful padding or serial processing.
Meanwhile, variable block sizes disrupt the uniform block-to-page mapping, forcing expensive KV gather operations before attention computation.
\sysname{} addresses these challenges with three dedicated GPU kernels.

\phm{Kernel 1: Fused query-centroid estimation.}
Since heads with different block sizes have varying numbers of centroids for the same context length, AB-Sparse stores all centroids in a flattened 1D layout and uses a prefix-sum indexing array to delimit the centroid segment of each head. 
Specifically, if head $h$ has $N_h$ centroids, we define $\mathrm{offset}_{h+1}=\mathrm{offset}_h+N_h$, so that the centroids of head $h$ are stored in $[\mathrm{offset}_h,\mathrm{offset}_{h+1})$.
This segmented layout enables all heads to be batched into a single kernel launch with fully vectorized execution and no padding overhead. 
We fuse dequantization into the kernel to avoid materializing dequantized centroids, reducing memory traffic.

\phm{Kernel 2: Batched Top-$K$ selection.}Given the estimation scores from kernel 1, this kernel selects the Top-$K_h$ blocks per head, where each head shares a fixed token budget $T$, and $K_h = \lceil T / B_h \rceil$ varies inversely with the assigned block size. 
This ensures that each head attends to the same number of tokens regardless of its block size, so that accuracy improvements stem from better block selection rather than increased token coverage.
The prefix-sum indexing array from kernel 1 is reused to partition the scores by head, avoiding redundant computation.

\phm{Kernel 3: Heterogeneous paged attention.}The final kernel computes attention over the selected blocks per head. 
The key challenge is that different heads have different block sizes, making standard paged attention kernels inapplicable without a costly gather step.
We avoid this by exploiting the hierarchical divisibility property between logical blocks and physical pages: any block naturally decomposes into an integer number of the finest-granularity pages.
As illustrated in \Cref{fig:solu_mapping}, each head's selected blocks are represented as a strided index view with no data movement, remaining fully compatible with existing paged attention kernels.

% As illustrated in \Cref{fig:solu_mapping}, kernel 3 maps the selected logical block indices from kernel 2 to their corresponding physical pages. Since the KV cache is managed in fixed-size physical pages, each logical block spans a contiguous range of physical pages determined by the block-to-page stride $s_h=B_h/P$, where $P$ is the physical page size. For each selected logical block index, the kernel expands it into $s_h$ consecutive physical page indices, producing the exact input format required by the standard paged attention kernel. This page mapping kernel serves as a precise connector between variable block size execution and standard paged attention, enabling seamless integration without any modification to existing inference infrastructure.
\section{Evaluation}
\label{sec:eval}

In this section, we perform quantitative experiments to demonstrate that \sysname{} improves accuracy over existing block sparse attention baselines while preserving throughput. We present accuracy results in \S\ref{sec:eval_acc}, efficiency results in \S\ref{sec:eval_eff}, micro study in \S\ref{sec:eval_micro}, and ablation studies in \S\ref{sec:eval_ablation}.

\subsection{Experimental Setup}

\phm{Hardware and models.}We conduct throughput experiments on two hardware platforms: NVIDIA A100-80GB and NVIDIA H800-80GB GPUs. We evaluate \sysname{} on three representative open-source LLMs: Llama-3.1-8B~\cite{llama-3.1-8B}, Qwen3-8B~\cite{qwen3}, and Qwen3-32B~\cite{qwen3-32b}, spanning two architecture families and natively supporting context lengths up to 128K tokens.

\phm{Benchmarks.}We employ two complementary benchmarks for accuracy evaluation: RULER~\cite{hsieh2024ruler} and LongBench~\cite{bai2024longbench}.
RULER is a synthetic benchmark designed to systematically probe long-context capabilities. 
It encompasses four task categories: retrieval, multi-hop reasoning, aggregation, and question answering, covering 13 tasks in total. 
We evaluate at context lengths from 16K to 96K to assess performance scaling with sequence length.
LongBench provides a more realistic evaluation suite comprising real-world long-document understanding tasks across six diverse categories: single-document QA, multi-document QA, summarization, few-shot learning, synthetic tasks, and code completion. This benchmark complements RULER by evaluating \sysname{} on natural text with practical downstream tasks.

\phm{Baselines.}We compare \sysname{} against full attention~\cite{dao2022flashattention} and two state-of-the-art block sparse attention methods: Quest~\cite{tang2024quest}~and ArkVale~\cite{chen2024arkvale}. 
Quest estimates block importance using per-block min-max pooling centroids, while ArkVale employs bounding-volume centroids for tighter block representation.
\sysname{} is applied on top of Quest and ArkVale as a drop-in replacement for their uniform block size assignment, with centroid quantization enabled.
For all sparse methods, we fix the KV budget at 4\% and the average block size at 32, following the settings adopted in common practice~\cite{liu2025freekv,wu2026prkvpage}. 

\subsection{Accuracy Evaluation}
\label{sec:eval_acc}

\begin{table*}[b]
    \centering
    \setlength{\tabcolsep}{1.8mm}
    \caption{Accuracy (\%) comparison on RULER (left) and LongBench (right) across three models. \sysname{} consistently outperforms baselines across all tasks.}
    \label{tab:eval_acc}
    \setlength{\tabcolsep}{6pt}
    \resizebox{1\columnwidth}{!}{
    \begin{tabular}{c|l|cccc|c|cccccc|c}
    \toprule
    & Methods & 16K & 32K & 64K & 96K & Avg. & SQA & MQA & SUM & FL & ST & CC & Avg. \\

    \midrule
    \multirow{5}{*}{\rotatebox{90}{\textit{Llama-3.1-8B}}} & Full Attention & 94.15 & 92.30 & 86.41 & 81.99 & 88.71 & 23.15 & 19.79 & 25.25 & 62.31 & 61.46 & 58.19 & 41.69 \\
    & Quest & 81.57 & 81.02 & 77.48 & 72.83 & 78.23 & 20.35 & 17.93 & 23.40 & 58.64 & 55.59 & 48.69 & 37.43 \\
    & \sysname{}-Quest & 85.04 & 85.56 & 81.49 & 74.87 & 81.74 & 22.19 & 18.03 & 25.04 & 60.85 & 60.24 & 53.05 & 39.90 \\
    & ArkVale & 82.77 & 83.50 & 80.69 & 75.43 & 80.60 & 22.40 & 18.11 & 23.57 & 59.21 & 56.88 & 53.18 & 38.89 \\
    & \sysname{}-ArkVale & 88.98 & 88.43 & 82.84 & 77.34 & 84.40 & 22.02 & 19.73 & 24.87 & 62.50 & 59.86 & 55.87 & 40.81 \\

    \midrule
    \multirow{5}{*}{\rotatebox{90}{\textit{Qwen3-8B}}} & Full Attention & 91.49 & 91.19 & 74.44 & 71.74 & 82.22 & 15.64 & 13.30 & 21.78 & 62.30 & 65.11 & 67.33 & 40.91 \\
    & Quest & 81.16 & 82.28 & 66.88 & 66.49 & 74.20 & 11.97 & 10.88 & 20.29 & 56.45 & 56.67 & 52.77 & 34.84 \\
    & \sysname{}-Quest & 84.00 & 84.01 & 69.50 & 68.03 & 76.39 & 13.68 & 12.13 & 20.86 & 59.09 & 61.30 & 57.61 & 37.45 \\
    & ArkVale & 84.51 & 84.52 & 68.99 & 67.90 & 76.48 & 13.35 & 11.93 & 20.34 & 57.18 & 57.26 & 58.37 & 36.41 \\
    & \sysname{}-ArkVale & 87.29 & 87.38 & 70.62 & 69.29 & 78.65 & 14.74 & 12.44 & 21.29 & 59.84 & 61.20 & 62.08 & 38.60 \\

    \midrule
    \multirow{5}{*}{\rotatebox{90}{\textit{Qwen3-32B}}} & Full Attention & 86.25 & 87.75 & 84.58 & 79.47 & 84.51 & 18.61 & 16.64 & 21.75 & 54.03 & 54.58 & 18.31 & 30.65 \\
    & Quest & 77.78 & 80.09 & 75.02 & 72.07 & 76.24 & 14.47 & 15.39 & 20.56 & 49.23 & 47.60 & 15.50 & 27.12 \\
    & \sysname{}-Quest & 86.98 & 84.86 & 79.34 & 75.50 & 81.67 & 16.78 & 15.87 & 20.72 & 52.36 & 56.27 & 16.43 & 29.74 \\
    & ArkVale & 84.84 & 84.79 & 79.19 & 74.25 & 80.77 & 15.56 & 15.60 & 20.19 & 49.89 & 52.79 & 15.46 & 28.25 \\
    & \sysname{}-ArkVale & 89.88 & 86.37 & 81.19 & 78.27 & 83.93 & 16.41 & 16.89 & 21.47 & 54.68 & 51.98 & 17.17 & 29.77 \\

    \bottomrule
    \end{tabular}
    }
\end{table*}

\Cref{tab:eval_acc} reports accuracy results on RULER (left) and LongBench (right). \sysname{} consistently outperforms both baselines across all models and benchmarks. \sysname{}-Quest improves over Quest by 3.51\%/2.19\%/5.43\% on RULER and 2.47\%/2.61\%/2.62\% on LongBench for Llama-3.1-8B/Qwen3-8B/Qwen3-32B, respectively. \sysname{}-ArkVale achieves similar gains, surpassing ArkVale by 3.80\%/2.17\%/3.16\% on RULER and 1.92\%/2.19\%/1.52\% on LongBench. These results confirm that adaptive block size allocation recovers a substantial fraction of the accuracy gap between sparse and full attention without increasing the average KV cache budget.

Notably, AB-Sparse consistently improves over both Quest and ArkVale, despite their fundamentally different block representation strategies. 
This suggests that adaptive block size allocation is agnostic to the underlying block representation, offering a general and pluggable enhancement for block sparse attention methods.

\subsection{Efficiency Evaluation}
\label{sec:eval_eff}
\begin{figure}[h]
    \centering
    \begin{minipage}{1\linewidth}
        \centering
        \includegraphics[width=1\linewidth]{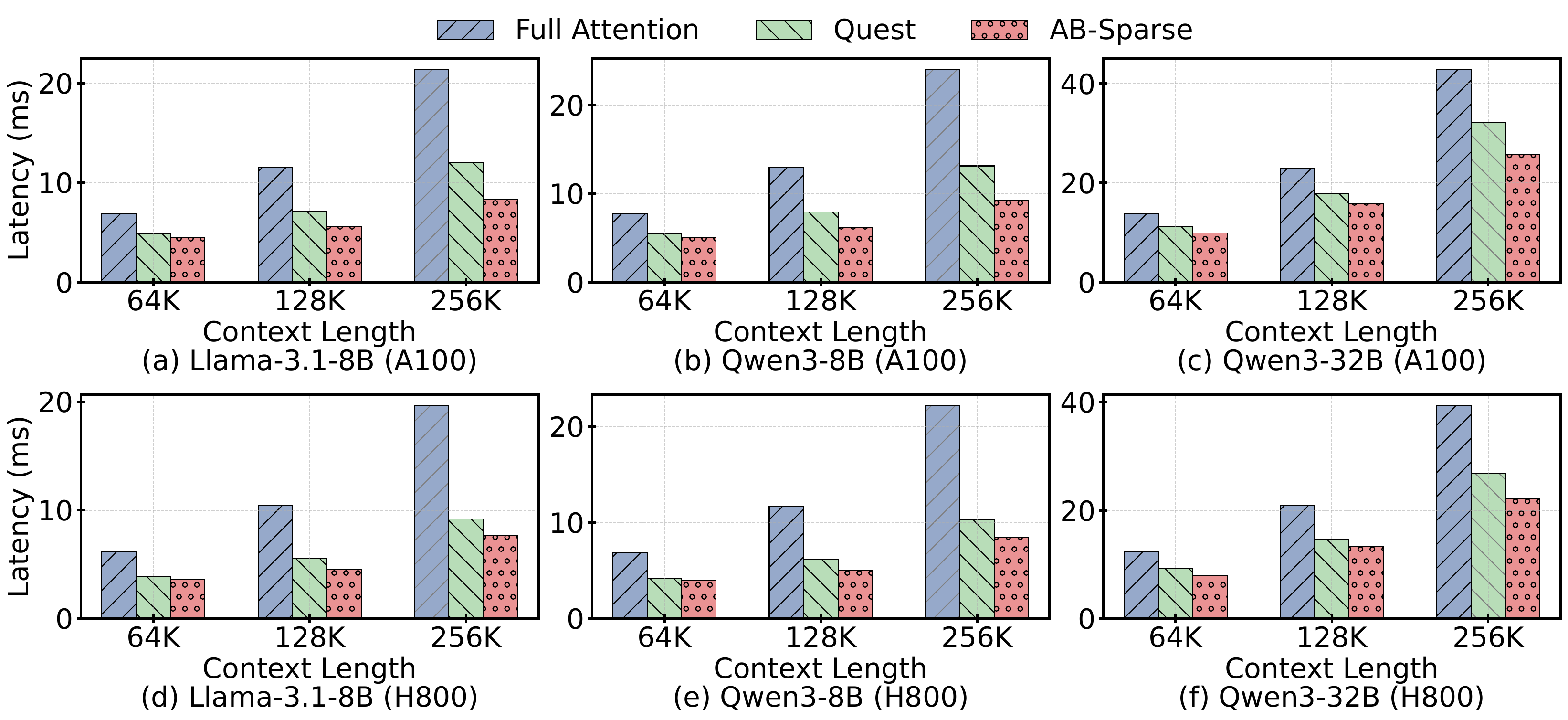}
        \caption{Decoding attention latency (ms) across three models with varying context lengths on A100 and H800 GPUs. \sysname{} achieves increasingly lower latency as context length grows.}
        \label{fig:eval_e2e}
    \end{minipage}
\end{figure}

We evaluate the decoding efficiency of \sysname{} against baselines across context lengths from 64K to 256K tokens. 
Since ArkVale differs from Quest only in block representation method, their latency characteristics are largely identical. 
We therefore exclude ArkVale from the efficiency comparison and use Quest as a representative block sparse attention baseline. \Cref{fig:eval_e2e} presents the decoding attention latency of all methods on A100 and H800 GPUs across all three models. \sysname{} matches Quest in latency at shorter contexts and becomes increasingly faster as context length grows. This is because INT4 centroid quantization reduces memory traffic during the estimation stage, an advantage that scales with context length.

% \begin{figure}[h]
%     \centering
%     \includegraphics[width=0.3\linewidth]{figure/abla_bsz.pdf}
%     \caption{Throughput (tokens/s) at \textbf{xxx}K context length with varying batch sizes (1--4) on Llama-3.1-8B.}
%     \label{fig:eval_bsz}
% \end{figure}

% \begin{table}[h]
% \centering
% \caption{Long generation accuracy (\%) (pass@4) on reasoning benchmarks using Qwen3-8B with 32K maximum generation length. \sysname{}-Quest consistently outperforms Quest.}
% \label{tab:eval_gene}
% \begin{tabular}{l|ccc|c}
% \toprule
% Methods & AIME24 & AMC23 & MATH-500 & Avg. \\
% \midrule
% Quest & 20.0 & 47.5 & 74.0 & 47.2 \\
% \sysname{}-Quest & 23.3 & 52.5 & 84.0 & 53.3 \\
% \bottomrule
% \end{tabular}
% \end{table}

% \begin{figure}[h]
%     \centering
%     \includegraphics[width=0.3\linewidth]{figure/abla_budget.pdf}
%     \caption{Throughput (tokens/s) at \textbf{xxx}K context length with varying batch sizes (1--4) on Llama-3.1-8B.}
%     \label{fig:eval_budget}
% \end{figure}

\begin{figure}[t]
\centering
\begin{minipage}[b]{0.3\textwidth}
    \centering
    \includegraphics[width=\linewidth]{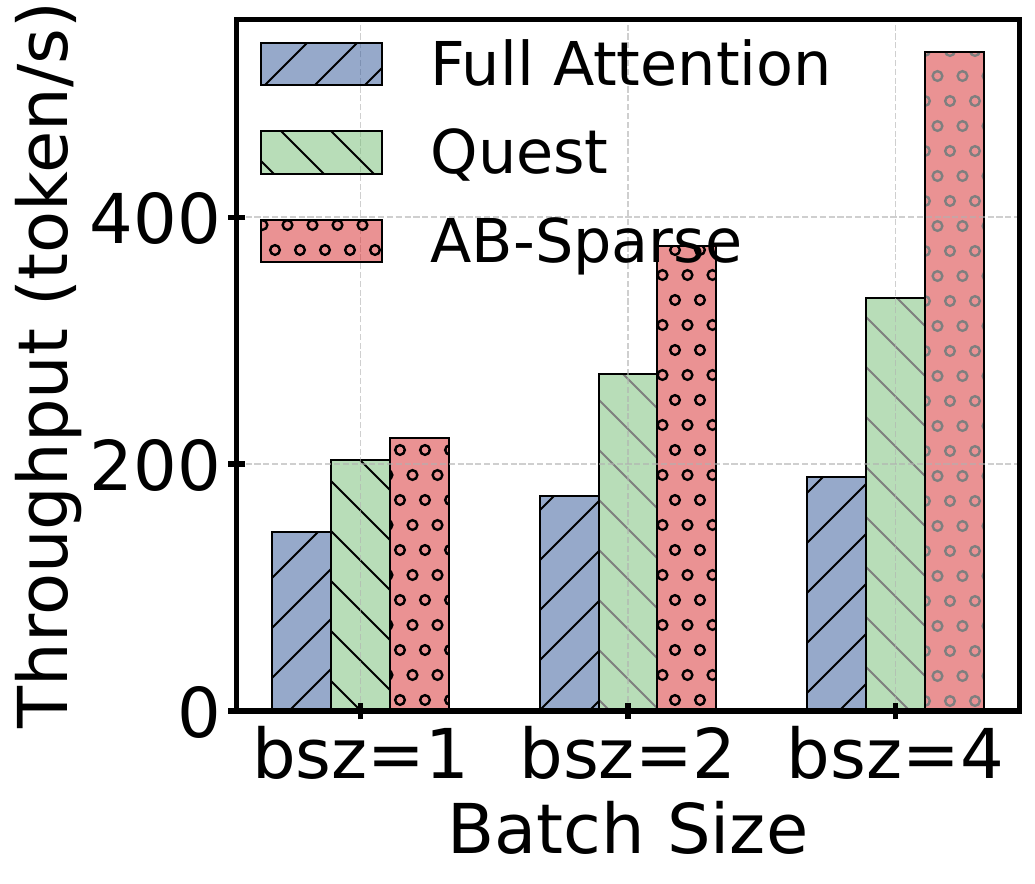}
    \captionof{figure}{Throughput (tokens/s) with 64K context length and varying batch sizes on Llama-3.1-8B.}
    \label{fig:eval_bsz}
\end{minipage}
\hfill
\begin{minipage}[b]{0.4\textwidth}
    \centering
    \small
    \begin{tabular}{l|cc}
    \toprule
     & Quest & \sysname{} \\
    \midrule
    AIME24  & 20.0 & 23.3 \\
    AMC23   & 47.5 & 60.0 \\
    MATH500 & 74.0 & 76.0 \\
    \midrule
    Avg.    & 47.2 & 53.1 \\
    \bottomrule
    \end{tabular}
    \vspace{2.4em} 
    \captionof{table}{Long generation accuracy (\%) (pass@4) of Qwen3-8B on three reasoning benchmarks with 32K max generation length.}
    \label{tab:eval_gene}
\end{minipage}
\hfill
\begin{minipage}[b]{0.25\textwidth}
    \centering
    \includegraphics[width=\linewidth]{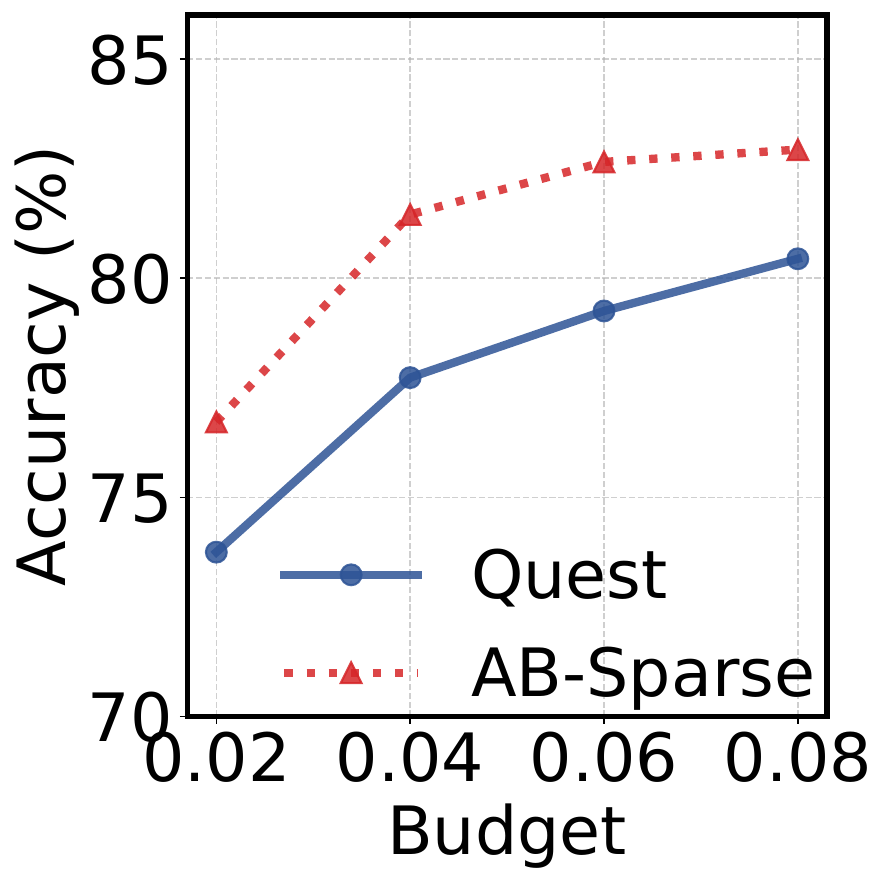}
    \captionof{figure}{RULER accuracy (\%) at 64K context length with varying token budget on Llama-3.1-8B.}
    \label{fig:eval_budget}
\end{minipage}
\end{figure}

\subsection{Microscopic Study}
\label{sec:eval_micro}
\phm{Throughput vs. batch size.}
\Cref{fig:eval_bsz} reports throughput of Llama-3.1-8B on A100 with 64K context length and batch sizes of $\{1,2,4\}$.
At batch size 1, \sysname{} achieves throughput comparable to Quest; at batch size 4, it reaches 1.59$\times$ the throughput of Quest.
This improvement stems from two factors: INT4 centroid quantization reduces memory traffic during the estimation stage, and the prefix-sum indexing enables padding-free batched execution across heads with heterogeneous centroid counts, allowing \sysname{} to scale more efficiently as batch size increases.

\phm{Long generation accuracy.}We additionally evaluate \sysname{} on long-generation tasks using Qwen3-8B~\cite{qwen3} on three reasoning benchmarks: AIME24~\cite{aime24}, AMC23~\cite{amc23}, and MATH-500~\cite{lightman2023letsverifystepstep}, which feature short inputs with long outputs. 
We adopt the sampling parameters recommended by Qwen3-8B~\cite{qwen3} (top\_k = 20, top\_p = 0.95, and temperature=0.6) and set the maximum generation length to 32K following DeepSeek-R1~\cite{guo2025deepseek}. 
We sample each input four times and report pass@4 as the accuracy metric. As shown in \Cref{tab:eval_gene}, \sysname{}-Quest outperforms Quest across three benchmarks, improving the average pass@4 from 47.2\% to 53.1\%. This demonstrates that adaptive block size allocation is effective not only for long-input tasks but also for long-generation tasks.

\phm{Dynamic token budget.}We evaluate Llama-3.1-8B on RULER at 64K context length, varying the token budget ratio from 2\% to 8\%.
As shown in~\Cref{fig:eval_budget}, \sysname{} consistently outperforms Quest across all budget levels by 2.97--3.69\%. 
The persistent gap as the budget increases suggests that adaptive block size allocation provides benefits complementary to simply enlarging the token budget.
Additional results on Qwen3-8B are provided in \S\ref{subsec:appendix_budget}.

\subsection{Ablation Study}
\label{sec:eval_ablation}

\begin{figure}[h]
    \centering
    \begin{minipage}[b]{0.47\linewidth}
        \centering
        \includegraphics[width=1\linewidth]{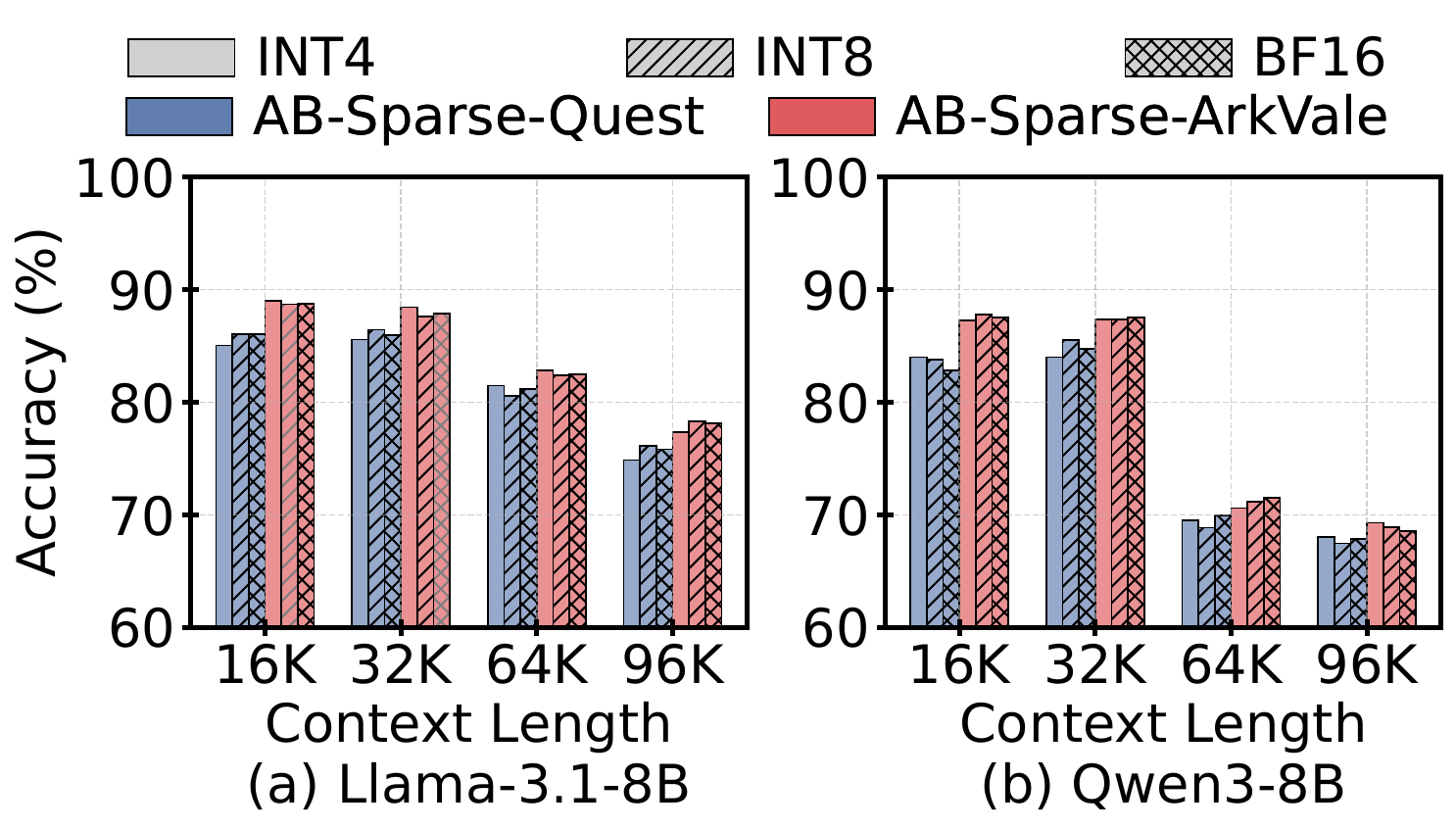}
        \caption{RULER accuracy (\%) under different centroid precisions across two models. INT4 quantization achieves accuracy comparable to the unquantized BF16 baseline.}
        \label{fig:eval_quant}
    \end{minipage}
    \hfill
    \begin{minipage}[b]{0.47\linewidth}
        \centering
        \includegraphics[width=1\linewidth]{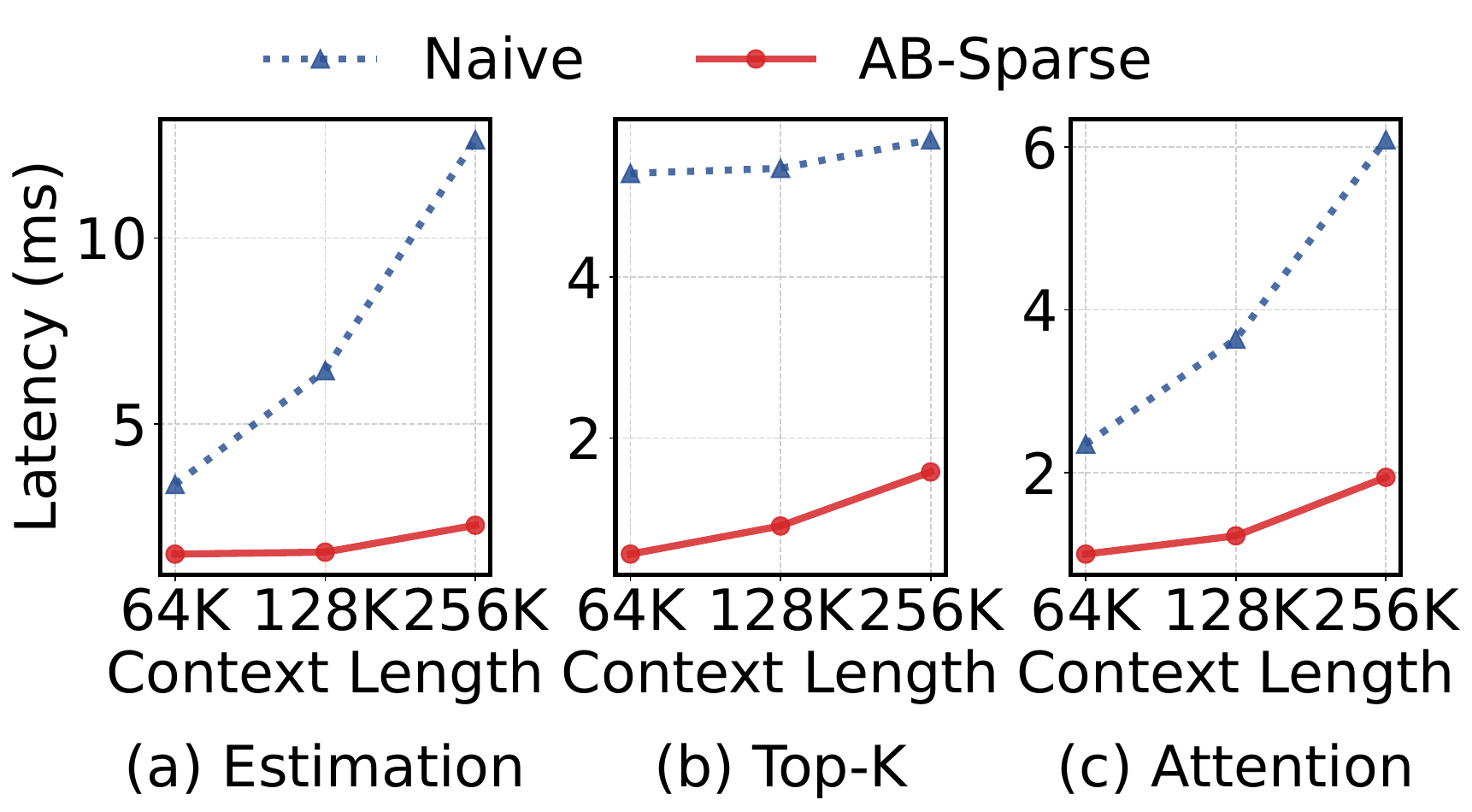}
        \caption{Kernel latency (ms) comparison between the naive implementation and \sysname{} with varying context length. AB-Sparse achieves consistently lower latency.}
        \label{fig:eval_kernel}
    \end{minipage}
\end{figure}

\phm{Effect of centroid quantization.}\Cref{fig:eval_quant} reports RULER accuracy under different centroid precisions across two models, with BF16 as the unquantized baseline.
INT4 quantization achieves accuracy comparable to BF16, confirming that per-channel asymmetric quantization preserves block ranking with negligible accuracy loss. Additional results on Qwen3-32B are provided in \S\ref{subsec:appendix_quant}.

% \begin{figure}[h]
%     \centering
%     \begin{minipage}{0.45\linewidth}
%         \centering
%     \includegraphics[width=1\linewidth]{figure/abla_quant_main.pdf}
%         \caption{RULER accuracy (\%) under different centroid precisions across two models. INT4 quantization achieves accuracy comparable to the unquantized BF16 baseline.}
%         \label{fig:eval_quant}
%     \end{minipage}
% \end{figure}

% \begin{figure}[h]
%     \centering
%     \begin{minipage}{0.45\linewidth}
%         \centering
%     \includegraphics[width=1\linewidth]{figure/abla_kernel.pdf}
%         \caption{RULER accuracy (\%) under different centroid precisions across two models. INT4 quantization achieves accuracy comparable to the unquantized BF16 baseline.}
%         \label{fig:eval_kernel}
%     \end{minipage}
% \end{figure}

\phm{Effect of custom kernels.}\Cref{fig:eval_kernel} compares the latency of the three core operations between the naive implementation and \sysname{}'s custom kernels across context lengths from 64K to 256K. 
The naive estimation and Top-$K$ kernels loop over heads sequentially due to varying centroid counts, while the naive attention kernel gathers selected KV blocks into contiguous memory before computation. 
Our kernels consistently achieve lower latency, with speedups of up to 5.6$\times$/9.4$\times$/3.1$\times$ for estimation/Top-$K$/attention, respectively.

\section{Conclusion}

We present \sysname{}, a training-free framework that improves the accuracy of block sparse attention by exploiting the heterogeneous block size sensitivity across attention heads. 
Through lightweight calibration-driven profiling, lossless centroid quantization, and efficient custom GPU kernels, \sysname{} achieves up to 5.43\% accuracy improvement on RULER and 2.62\% on LongBench over existing baselines, without throughput overhead.

% \input{8-ack}

% \newpage
\bibliographystyle{unsrt}
\bibliography{ref}

% %%%%%%%%%%%%%%%%%%%%%%%%%%%%%%%%%%%%%%%%%%%%%%%%%%%%%%%%%%%%
% \input{9-appendix}
% %%%%%%%%%%%%%%%%%%%%%%%%%%%%%%%%%%%%%%%%%%%%%%%%%%%%%%%%%%%%

% \newpage
% \input{10-checklist}

\end{document}